# Fast Uncoiling Kinetics of F1C Pili Expressed by Uropathogenic *Escherichia coli* are Revealed on a Single Pilus Level using Force-Measuring Optical Tweezers

Running title: Biomechanical properties of F1C pili

Mickaël Castelain,[1,3,5] Sarah Ehlers,[1,3] Jeanna Klinth,[1,3] Stina Lindberg,[2,3,4] Magnus Andersson,[1,3] Bernt Eric Uhlin,[2,3,4] and Ove Axner[1,3,6]

[1]*Department of Physics,* [2]*Department of Molecular Biology,* [3]*Umeå Centre for Microbial Research (UCMR), and* [4]*The Laboratory for Molecular Infection Medicine Sweden (MIMS), Umeå University, SE-901 87 Umeå, Sweden*

[5]*Present address: Laboratoire d'Ingénierie des Systèmes Biologiques et Procédés, INRA UMR792 – CNRS UMR5504, INSA F-31077 Toulouse Cedex 4, France*

[6]*Corresponding author:* ove.axner@physics.umu.se

## Abstract

Uropathogenic *Escherichia coli* (UPEC) express various kinds of organelles, so-called pili or fimbriae, that mediate adhesion to host tissue in the urinary tract through specific receptor-adhesin interactions. The biomechanical properties of these pili have been considered important for the ability of bacteria to withstand shear forces from rinsing urine flows. Force measuring optical tweezers have been used to characterize individual organelles of F1C type expressed by UPEC bacteria with respect to such properties. Qualitatively, the force-*vs.*-elongation response was found to be similar to that of other types of helix-like pili expressed by UPEC, i.e. type 1, P, and S, with force-induced elongation in three regions of which one represents the important uncoiling mechanism of the helix-like quaternary structure. Quantitatively, the steady-state uncoiling force was assessed to 26.4 ($\pm$ 1.4) pN, which is similar to those of other pili (which range from 21 pN for $S_I$ to 30 pN for type 1). The corner velocity for dynamic response (1400 nm/s) was found to be larger than those of the other pili (400 -700 nm/s for S and P pili, and 6 nm/s for type 1). The kinetics were found to be faster, with a thermal opening rate of 17 Hz, a few times higher than S and P pili, and three orders of magnitude times higher than type 1. These data suggest that F1C pili are, like P and S pili, evolutionary-selected to primarily withstand the conditions expressed in the upper urinary tract.

## 1. Introduction

The widespread bacterial resistance to antibiotics is a rapidly growing problem and there is therefore an urgent need for new anti-microbial drugs that can combat bacterial infections, particularly those caused by antibiotic-resistant bacteria. This necessitates the identification of new targets in bacteria which, in turn, requires detailed knowledge of microbial pathogenic mechanisms. Since adhesion of bacterial pathogens to host tissue is a prerequisite for infections, the adhesion mechanism is one such possible target.

Extraintestinal pathogenic *Escherichia coli* (ExPEC) are known to express, on the surface of the bacterial cells, a variety of flexible adhesion organelles, referred to as pili or fimbria. Uropathogenic *E. coli* (UPEC) is a type of ExPEC that is commonly associated with community–acquired urinary tract infections and expresses pili that are ~1 μm long, 6-7 nm in diameter, and consist of a large number ($>10^3$) of a given type of fimbrial subunit, assembled via the chaperone-usher pathway (Sauer et al. 2000; Sauer et al. 2004), in a head-to-tail-manner by a donor-strand-exchange mechanism (Poole et al. 2007; Sauer et al. 1999), which makes up the backbone of the pili. The subunits are ordered in a right-handed helix-like quaternary structure, referred to as the rod, where one layer is connected to an adjacent layer via layer-to-layer (LL) interactions with ~3 subunits per turn (Bullitt and Makowski 1995; Jass et al. 2004; Mu and Bullitt 2006). At the end of the helix-like rods, a short thinner protein polymer (a tip fibrillum) is present, which includes the adhesin protein that binds to specific receptor molecules that are part of glycolipids or glycoproteins on the host cells (Jones et al. 1995).

A bacterium needs to remain attached to host tissue to be able to colonize efficiently. Prompted by the fact that an ascending bacterium is exposed to significant forces from rinsing urine flows, recent studies have shown that the adhesive potential of bacteria is not solely given by the properties of the adhesin molecule located at the distal end of the adhesion organelles; it can be strongly influenced also by the biomechanical properties of the pilus shaft (Duncan et al. 2005). The quaternary structure of the pili expressed by UPEC can allow for both a redistribution of external force to several pili and provide a dampening effect, which together weaken or optimize the load applied to a given adhesin-receptor binding (Björnham and Axner 2009; Forero et al. 2006). Although the three dimensional static structure of only P pili has been assessed in some detail (Bullitt and Makowski 1995; Mu and Bullitt 2006; Verger et al. 2007), the biomechanical properties of the most common types of pili expressed by UPEC, P pili (pyelonephritis-associated pili), type 1, and S pili, have been characterized (Andersson et al. 2007; Castelain et al. 2009b; Fällman et al. 2005). It was found that their force-*vs.*-elongation response could be differentiated into three distinct extension regions, referred to as region I, II, and III (Andersson et al. 2006c; Jass et al. 2004). Region I shows a linear increase in force that indicates an elastic stretching of the pilus. The second region (II) displays a constant force-extension response, which originates from a successive unzipping of consecutive layers of the helix-like structure (below referred to as uncoiling, also denoted unfolding in the literature), which in turn leads to a linearization of the quaternary structure of pili. The third region (III) is characterized by a wave-like pseudo-elastic response, which originates from a conformational change in the head-to-tail interaction that takes place in a randomized order. Counter-intuitively, this elongation has been found to be reversible (Fällman et al. 2005), which implies that the linearized quaternary structure of pili can be recoiled (refolded) to its original helix-like configuration a repeated number of times without any fatigue (Andersson et al. 2006b). Since region II allows for a significant elongation of the pili length (5 – 7 times), it has been considered of special importance for the ability of a bacterium to redistribute an external shear force among a multitude of pili (Björnham and Axner 2009). This force redistribution is assumed to play a significant role for the ability of individual bacterial cells to sustain the strong shear forces to which they are exposed by rinsing urinary flow. This efficient machinery opens up for a possibility to reduce infections by UPEC bacteria by interfering with the adhesion mechanism, e.g. by small molecules that can compromise either the biogenesis of the biomechanical properties of helix-like pili (Åberg and Almqvist 2007; Åberg et al. 2007) or the recoiling process. However, to succeed in this, the mechanisms leading to the unique elongation and retraction properties of the pili need to be assessed on a detailed level, which in turn requires a characterization of the biomechanical properties of the various pili on a single-pilus level.

Although the most common types of UPEC pili, i.e., P, type 1, and S, are genetically and structurally similar, assembled via the chaperon-usher pathway and sharing a common overall helix-like architecture, it has been found that they have dissimilar biomechanical properties. Their type-specific properties are assumed to have evolved as a result of disparate environmental conditions and are therefore believed to be of specific importance for adhesion of multipili bacteria in different parts of the urinary tract. F1C, which likewise is expressed by UPEC bacteria and have the same general helix-like structure, recognize and attach to kidney epithelial (distal tubules and collecting ducts) and endothelial (bladder and kidney) cells and have been found to be encoded by ~14% of UPEC isolates. However, although they are believed to have an impact in urinary tract infection cases (ADD Khan et al. 2000, Ott et al., 1988; Hacker et al., 1993), little is known about their biomechanical properties. To increase the knowledge about F1C and its role in urinary tract infections, this paper reports on an assessment of these properties of F1C pili on a single-molecule level.

Since the Force Measuring Optical Tweezers (FMOT) technique repeatedly has been shown to be suitable for assessment of the biomechanical properties of pili, both those acting as molecular motors, e.g. type IV, (Allemand and Maier 2009; Biais et al. 2010; Merz et al. 2000) and those that are passive, primarily those with a helix-like structure, e.g. P, type 1, and S (Andersson et al. 2006c; Andersson et al. 2007; Castelain et al. 2009b; Fällman et al. 2005; Jass et al. 2004) but also those with a coil-like geometry, T4, (Castelain et al. 2009a) it has been used also for this study. The technique was used for both ordinary force-*vs.*-elongation studies, under steady-state as well as dynamic conditions, and relaxation investigations. By use of these methodologies, the main characteristic biomechanical properties of the F1C pilus, primarily in the two regions containing conformational changes of the pili, i.e. region II and to certain extent also region III, could be assessed on an individual pili level (Andersson et al. 2008; Axner et al. 2009). From assessments of various measurable entities [the uncoiling (unfolding) force in region II under steady-state and dynamic conditions, the shape of region III, and the relaxation time] and by the use of previously developed theoretical descriptions of pili elongation and retraction (Andersson et al. 2006a, c; Axner et al. 2009; Fällman et al. 2005; Jass et al. 2004), a variety of bond parameters can be assessed (e.g. the bond length and the bond opening length, the bond energy, the activation energy, and the thermal uncoiling rate of the important LL bond). These are then compared with the corresponding parameters assessed for the other major types of pili expressed by UPEC, i.e. P (Andersson et al. 2006c; Fällman et al. 2005; Jass et al. 2004), type 1 pili (Andersson et al. 2007) and S pili (Castelain et al. 2009b). It is found that although the steady-state behavior of F1C pili is similar to that of most other pili, the dynamic and temporal properties are different; they have faster kinetics than the others, which in turn indicates that they can temporarily sustain high forces.

## 2. Material and methods

### 2.1. Bacterial strain and sample preparation

A laboratory strain of *E. coli* that lacks expression of genes for pilus biogenesis (HB101) was used as a host for plasmids harboring fimbrial gene clusters cloned from UPEC isolates J96 (Hull et al. 1981). F1C pili addressed in this work were expressed from the plasmid that carries the *foc* gene cluster (HB101/pBSN50) (Lindberg et al. 2008; Lund et al. 1988). The strains were grown at 37°C overnight on 1.5% trypticase soy agar (Becton, Dickinson and Co.) complemented with carbenicillin (Duchefa Biochemie, 50 μg/ml).

Samples of bacteria expressing pili were harvested from the agar plate and suspended in a phosphate buffered saline solution (PBS (Sigma-Aldrich), 10 mM, 130 mM NaCl, pH = 7.4 at 25°C) together with 3 μm polystyrene beads (Duke Scientific Corp., Palo Alto, CA). Twenty-five μL of this suspension was then placed between two coverslips. Beforehand, large 9.6 μm beads (Duke Scientific Corp., Palo Alto, CA) were immobilized to the surface of the lower coverslip through heating at 60°C for 60 minutes. The large beads were treated with poly-L-Lysine (Sigma-Aldrich) in order to provide a positively–charged surface for attachment of single negatively–charged bacterial cells. This creates strong electrostatic bonds with the bacteria, substantially stronger than the bead–pili interaction, which ensures that the bacterium is properly fixed during an experiment.

### 2.2. Atomic Force Microscopy (AFM)

F1C pili expressed by HB101/pBSN50 cells were imaged by atomic force microscopy (AFM) as described earlier (Balsalobre et al. 2003). The final image was flattened and plane fitted in both axes using Veeco software and presented in height mode.

*2.3. Optical tweezers – Mounting and measurement procedures*

Following the transfer of the mixture of bacteria and small beads (with a diameter of 3 µm, serving as the handle for the optical tweezers) to the cover glass, bacteria were gently captured by optical tweezers and mounted onto the large beads. Force measuring optical tweezers were used to extend individual pili under both steady-state and dynamic conditions according to a procedure previously used for characterization of other types of pili (Andersson et al. 2006a, c; Andersson et al. 2007; Castelain et al. 2009b; Fällman et al. 2004; Jass et al. 2004). In short, a small bead was trapped and brought to a position close to the bacterium. The trap was then calibrated with the power spectrum method (Fällman et al. 2004). The trap stiffness was typically 180 pN/µm. The probe bead was thereafter moved back and forth close to the bacterium until pili attachment took place. Since the probe beads were not functionalized with specific receptors, the pili bound non-specifically to the bead. Because of this, not only the tip fibrillum, but possibly also a part of the rod, might have been attached the probe bead. The part of the pili being elongated might therefore not be identical to the full length of the rod.

The data acquisition was started and the piezo-stage set in motion to separate the bacterium from the small bead. As is further illustrated below, it was generally found that the bead was attached to the bacterium by several pili. However, as the separation increased, the pili detached from the bead one at the time. The separation continued until a single pilus remained. The stage was then stopped and the motion reversed while the remaining pilus was still attached. When the remaining pilus had been contracted to its original length, the motion was again halted. The two halt positions were then defined as the start and a stop position for repetitive elongations. The remaining pilus was then elongated and allowed to retract repeatedly between these positions with a given elongation speed by the use of a custom made computer program that controlled the traveling and the speed at which the stage was moved. Each pilus could thereby be exposed to a repeated number of extension/contraction cycles, under controlled and identical conditions, reducing the statistical fluctuations in the assessment of the force-*vs.*-elongation response.

The force response of the pili was assessed by ordinary force-*vs.*-elongation procedures for a variety of elongation speeds, ranging from 0.1 to 243 µm/s, which encompass assessments under both steady-state and dynamic conditions, as well as by relaxation investigations. The assessment under steady state conditions was based upon around 230 measurements, whereas those under dynamic conditions were based on fewer, typically ~15 per elongation speed. The relaxation measurements consisted of an elongation of the pilus at a constant speed under dynamic conditions (i.e. with a rather high elongation speed, typically 40 µm/s) after which the elongation was suddenly halted and the decay of the uncoiling force (from its dynamic to its steady-state level) was measured.

## 3. Theory

The fact that the main part of the pili expressed by UPEC bacteria constitutes a rod that is assembled from a number of subunits arranged in a helix-like quaternary structure provides the pili with a unique and rather intricate force-*vs.*-elongation behavior. Extensive descriptions of the basic features giving rise to this behavior have been given previously in the literature (Andersson et al. 2006c; Axner et al. 2009). To facilitate the interpretation of the data taken and to provide a useful nomenclature for the analysis, a brief summary is given here.

The rod consists of a large number of repeated subunits connected by hydrophobic interactions (Mu and Bullitt 2006), here simply referred to as head-to-tail (HT) bonds, which constitute the tertiary structure and make up the backbone of the pilus. The quaternary structure is formed by means of LL bonds that connect non–consecutive subunits to each other, predominantly the subunits $n$ and $n+3$, which gives rise to a helix-like structure comprising slightly more than three LL-bonds per turn. When exposed to stress, the rod will uncoil in a sequential manner [uncoiling of the rod at an arbitrary position in the interior of the rod is improbable since it requires the opening of a multitude of bonds simultaneously (Andersson et al. 2008)]. The elongation rate of the pilus in region II is therefore simply given by the net opening rate of the outermost not-yet opened LL bond times the bond opening length of the LL bond.

The opening and closure behavior of a single bond is well described by rate theories based on the works by Kramers (Kramers 1940) and Bell (Bell 1978). The behavior of the entire pilus, when exposed to stress or strain (i.e. force or elongation), has been implemented both in a sticky-chain model (Andersson et al. 2006a, c) and by Monte Carlo simulations (Björnham et al. 2008) and both predict force-$vs.$-elongation behaviors that agree well with the experimental findings of the types of pili so far assessed, P (Fällman et al. 2005; Jass et al. 2004), type 1 (Andersson et al. 2007), and S (Castelain et al. 2009b). The same type of description is therefore used to analyze the force-$vs.$-elongation behavior of F1C pili studied in this work.

### 3.1. Region II

In short, the rate equation for the number of open LL bonds in the rod, $N_B$, exposed to a force, $F$, which thus also gives the net opening rate of the outermost bond in the rod, $dN_B/dt$, is given by

$$\frac{dN_B}{dt} = k_{AB}^{th} e^{F\Delta x_{AT}/kT} - k_{AB}^{th} e^{(\Delta V_{AB} - F\Delta x_{TB})/kT} \,, \qquad (1)$$

where $k_{AB}^{th}$ is the thermal bond opening rate, $\Delta x_{AT}$ and $\Delta x_{TB}$ are the bond lengths for opening and closure of the bond, i.e., the distance from the closed and open state to the transition barrier, respectively, and $\Delta V_{AB}$ is the bond energy for the LL bond. $k$ is the Boltzmann factor and $T$ the temperature. At room temperature, the $kT$ product takes a value of 4.1 pN·nm.

For low elongation speeds, i.e. for $\dot{L} < \dot{L}^{*}$, where $\dot{L}$ is the elongation speed of the pilus and $\dot{L}^{*}$ an entity referred to as the corner velocity, defined below (Andersson et al. 2006a), which often is referred to as **steady-state conditions**, the opening and closure rates are in balance, which leads to a steady-state uncoiling force, $F_{UC}^{SS}$, given by

$$F_{UC}^{SS} = F_{UC}(\dot{L} < \dot{L}^{*}) = \frac{\Delta V_{AB}}{\Delta x_{AB}}, \tag{2}$$

where $\Delta x_{AB}$ is the bond opening length of the LL bond, given by $\Delta x_{AT} + \Delta x_{TB}$ (Andersson et al. 2006c).

For high extension speeds, i.e. for $\dot{L} > \dot{L}^{*}$ (Andersson et al. 2006a; Axner et al. 2009), referred to as **dynamic conditions**, the refolding rate in Eq. (1) can be neglected and the uncoiling force, denoted by $F_{UC}^{D}$, depends on the elongation speed in a logarithmic manner, i.e. as

$$F_{UC}^{D} = F_{UC}(\dot{L} > \dot{L}^{*}) = \frac{kT}{\Delta x_{AT}} \ln(\frac{\dot{L}}{\dot{L}^{th}}), \tag{3}$$

where $\dot{L}^{th}$ is the thermal extension speed, given by $k_{AB}^{th} \Delta x_{AB}$ (Andersson et al. 2006a). The corner velocity is the velocity for which Eq. (3) predicts an uncoiling force equal to $F_{UC}^{SS}$, i.e.

$$\dot{L}^* = \dot{L}^{\text{th}} e^{F_{UC}^{SS} \Delta x_{\text{AT}}/kT} .$$  (4)

It is customary to plot the uncoiling force versus the logarithm of the elongation velocity and fit the Eqs (2) and (3) to the low and high velocity parts of the curve, respectively. Equation (3) implies that a pilus exposed to a force above the steady-state unfolding force will elongate at a velocity given by

$$\dot{L}(F) = \dot{L}^{\text{th}} e^{F \Delta x_{\text{AT}}/kT} ,$$  (5)

which necessarily is larger than the corner velocity. Investigations under dynamic conditions, which comprise assessments of the uncoiling force as a function of elongation speed, are usually referred to as dynamic force spectroscopy (DFS) (Andersson et al. 2006a).

The **temporal behavior** of a pilus can be assessed by so called relaxation investigations (Andersson et al. 2007), in which the pilus first is elongated at a speed that provides a dynamic response before the elongation is suddenly halted. Under these conditions, the uncoiling force will experience a time-dependence that is given by

$$\frac{dF_{\text{UC}}(t)}{dt} = -\dot{L}^{\text{th}} \kappa \left[ e^{F_{\text{UC}}(t) \Delta x_{\text{AT}}/kT} - e^{(\Delta V_{\text{AB}} - F_{\text{UC}}(t) \Delta x_{\text{TB}})/kT} \right],$$  (6)

where $\kappa$ is the elastic constant of the force transducer (Andersson et al. 2007). Investigations under dynamic and temporal conditions provide information related to the transition state that is not addressable under steady-state conditions, primarily the bond length, $\Delta x_{\text{AT}}$, and the energy of the transition state, $\Delta V_{\text{AT}}$.

### 3.2. Region III

As was alluded to above, when the entire helix–like quaternary structure has been linearized, the HT–bonds will, during continued pili–elongation, rapidly take up a larger force, whereby they will be subjected to conformational changes, which gives rise to a wave–shaped force–*vs.*–elongation dependence, referred to as region III. Since the HT bonds in a linearized pilus can alter their conformational state in a random manner, the particular shape of this region is governed by both properties of the individual bonds and entropy, of which the latter gives it its specific soft, wave-like shape. It has previously been shown that under steady-state conditions the force response of region III can be written as

$$F_{\text{III}}^{\text{SS}}\left(N_{\text{B}}\right) = \frac{\Delta V_{\text{BC}}}{\Delta x_{\text{BC}}} + \frac{kT}{\Delta x_{\text{BC}}} \ln\left(\frac{N_{\text{TOT}} - N_{\text{B}}}{N_{\text{B}}}\right), \tag{7}$$

where $\Delta V_{\text{BC}}$ and $\Delta x_{\text{BC}}$ are the differences in energy between an open and closed head-to-tail bond (in the absence of stress) and the bond opening length, respectively, and $N_{\text{TOT}}$ and $N_{\text{B}}$ are the total number of units in the rod and the number of closed head-to-tail bonds in the linearized pili, respectively. Since the length of the pilus, $L$, can be related to $N_{\text{TOT}}$ and $N_{\text{B}}$ by geometrical means, Eq. (7) provides an expression for the force-*vs.*-elongation (as well as the force-*vs.*-contraction) behavior of pili in region III under steady-state conditions (Andersson et al. 2006c, d).

Since the HT-bonds can open in a random manner, the corner velocity of region III is orders of magnitude above that of region II. This implies that the elongation/retraction in region III, under all practical conditions, takes place under steady-state conditions, which in turn indicates that some energy landscape parameters for the head-to-tail bond, in particular those involving the transition state between state B and C, e.g. $\Delta x_{BT}$ and $\Delta V_{BT}$, can neither readily be assessed, nor are they expected to play any role for the biomechanical properties of the pili.

## 4. Results

### 4.1. Imaging

Figure 1, which provides an AFM micrograph of HB101/pBSN50 cells expressing F1C pili, shows that the bacterial strain used expresses a multitude of pili of a variety of lengths, predominantly up to a couple of micrometers.

**Figure 1 here**

### 4.2. Characterization of F1C pili under steady-state conditions

The initial bead-bacterium interaction provides in general multipili adhesion. Such adhesion will limit the amount of information that can be retrieved from the system since it will not give an unambiguous picture of the fundamental interactions in the system, primarily due to a partly unknown interplay between the various pili. On the other hand, as is shown in figure 2a, which shows one-and-a-half elongation-retraction cycle of a typical force-extension/contraction response of F1C pili under steady-state conditions, single pilus responses can be obtained by elongating a multi-pili response until only one pilus remains attached.

The uppermost curve in figure 2a (red on-line) displays the typical force response from the first elongation cycle of a bacterium-bead interaction, thus showing the response from a multipili attachment. As the pili are elongated, they successively detach, marked by arrows, until only one pilus remains. As is further discussed below, single pilus response takes place, in this particular case, from 5.2 µm (from which the remaining pilus is elongating in region II, until 6.5 µm when it enters region III). After 12 µm, the elongation is halted. By then retracting the piezo-stage to the starting position (shown by the grey curve), the single pilus will retract and recoil. When the remaining pilus is no longer elongated (i.e. when it has reached region I), the retraction is again stopped. By then again moving the piezo-stage in the positive direction, the remaining single pilus will be elongated. Repeated single pilus measurements on the same pilus (both elongation and retraction) can then be performed a multitude of times (of which the first elongation is shown by the black curve in the figure).

**Figure 2 here**

As can be seen from the black curve in figure 2a, a single helix-like pilus has an intrinsic force response that differs from that of a single bond as well as from those of many other types of biopolymer. As was briefly mentioned above, the force-*vs.*-elongation response can be seen as composed of three regions; **Region I**, in which the response is basically linear, like that of a normal (elastic) spring; **Region II**, in which the force response is constant (i.e. elongation-independent), seemingly like that of a material that undergoes plastic deformation; and **Region III**, in which the response has a monotonically increasing but non-linear force-*vs.*-elongation response.

Whereas the first region, in which the pilus is elongated a fraction of its length, can be understood as a general (elastic) stretching of the pilus, but with no conformational change (no opening or closure of bonds), the other regions originate from an opening, a closure, or a conformational change of the individual non-covalent bonds (or interactions) that connect the various subunits in the pilus. The constant force-*vs.*-elongation response in region II, which in this case takes place at a force of ~26 pN, is a direct result of an uncoiling of the helix-like quaternary structure of the pilus by a sequential opening of the LL bonds. As was mentioned above, the reason the opening of bonds is sequential in this region is that each turn of the quaternary structure of the pilus consists of several subunits (slightly more than three, depending on type of subunit). Since each subunit can mediate one LL bond, there are ~ three LL bonds per turn. When a pilus is exposed to a force, each bond in the interior of the rod will therefore experience approximately only one third of the applied force. The bond connecting the outermost unit in the folded part of the rod, on the other hand, experiences a significantly higher force, virtually the entire force to which the pilus is exposed. This implies that the outermost LL bond of the rod will open significantly more often than a bond in the interior (for an uncoiling process in the interior of the rod to occur, three successive LL bonds, each exposed to about a third of the entire force, have to open simultaneously, which is an exceedingly rare process). This implies, in turn, that the uncoiling of the rod takes place predominantly by a sequential opening of the outermost LL bond, sometimes referred to as a zipper-like uncoiling, which takes place at a constant force and thus shows up as a force plateau. The plateau is truly flat and remarkably repeatable; its value does not spread more than a few percent between consecutive measurements on the same pilus and slightly more between different individual pili, which is considered to originate mainly from fluctuations in the instrumentation conditions, e.g. the bacterial mounting geometry. Figure 2b gives a histogram over the distribution of the uncoiling force in region II, encompassing 230 measurements performed under steady-state conditions. The data provide an uncoiling force of F1C pili under steady-state conditions of 26.4 ± 1.4 pN.

In contrast, the soft wave-like force-*vs.*-elongation behavior (a pseudo-elastic compliance) in region III (black curve in figure 2a) originates from a stochastic conformational change of the HT-bonds between consecutive subunits of the pilus in the linearized part of the rod. Since these bonds can alter their conformational state in a random order, the particular shape of this region is governed by both properties of the individual bonds and entropy, of which the latter gives it its specific wave-like shape. A detailed theoretical analysis of region III has previously been given (Andersson et al. 2006c). It was shown that by fitting a curve of the form given by Eq. (7) to the data, or, alternatively, by identifying three slopes in the curve, it is possible to predict the number of subunits involved in the elongation process of a given pilus (Andersson et al. 2006d), $N_{TOT}$, which in turn can be used to assess the bond opening length $\Delta x_{AB}$ from the elongation of region II. Based upon 16 curves, $\Delta x_{AB}$ was assessed to $5.6 \pm 0.4$ nm. Moreover, as is shown by Eq. (2), a measurement of the unfolding force allows for an assessment of the bond energy of the LL-bond, $\Delta V_{AB}$, which for F1C was found to be $36 \pm 2 \ kT$.

In addition, a comparison of the grey and the black curves in figure 2a, representing retraction and elongation of a single pilus, respectively, illustrates that for a elongation/retraction velocity of 0.1 µm/s the response is more or less fully reversible, which concur with previous findings of P pili (Fällman et al. 2005; Jass et al. 2004). This illustrates the interesting phenomenon that despite the fact that the elongation in region II takes place at a constant force, it does not originate from a plastic deformation, as is customary for solid materials; it originates from a fully ordered and repeatable uncoiling/recoiling process.

It should be noticed though that during contraction (the grey curve), following the transition from region III to II, a dip is observed. This dip is fully reproducible and has previously appeared for both P and type 1 pili (Fällman et al. 2005; Lugmaier et al. 2008). The decrease in force is considered to be caused by a lack of a nucleation kernel for the formation of a first layer in the quaternary structure of the pilus during contraction (after being fully linearized, the pilus needs a certain amount of slack to form the first turns of the LL interactions). After formation of the first part of the helix-like structure, at which the force returns to the ordinary refolding force, the refolding continues in a sequential manner until all subunits are stacked into the original helix-like shape. Occasionally, also other dips in the contraction data can occur, as for example is the case for the data shown in figure 2a at around 1 μm, which are assumed to originate from sporadic misfoldings.

From an understanding of the single-pilus response, the multipili response illustrated by the uppermost (red on-line) curve in figure 2a can now be partly understood. Based upon the elongation-independent uncoiling force of region II, which was found to be $26.4 \pm 1.4$ pN, the two force plateaus at ~ 75 pN (from 2 to 3.2 μm, denoted region $II^{(3)}$) and at ~ 50 pN (from 3.5 to 5 μm, region $II^{(2)}$) can be attributed to the simultaneous uncoiling of 3 and 2 pili, respectively. The first part of the elongation, up to 1.7 μm, is more complex, presumably involving the elongation of one or several pili in region I.

### 4.3. Characterization of F1C pili under dynamic conditions

As was alluded to above, although the uncoiling forces of the various types of pili are virtually velocity independent at low elongation speeds (below the corner velocity $\overset{*}{L}$, for which the pili are uncoiling under steady-state conditions), for high elongation speeds (above the corner velocity) the uncoiling takes place under dynamic conditions (Andersson et al. 2006a), in which the uncoiling force is excepted to increase with elongation velocity (in a logarithmic manner).

Elongating the F1C pili in region II at different speeds gives indeed rise to a variety of uncoiling force levels. Figure 3a shows, by the black curve, an individual F1C pilus elongated at a low velocity (at 0.1 µm/s, i.e. under steady-state conditions, with the corresponding retraction in grey) together with parts of the response from elongations at higher velocities, at 3, 9 and 64 µm/s, which demonstrate elongation in the dynamic regime. It can clearly be seen that the uncoiling response in region II is affected by the elongation velocity although the response in region III is independent, in agreement with the predictions above.

**Figure 3 here**

Figure 3b displays, in a semi-logarithmic plot, the uncoiling force in region II as a function of elongation speed. The individual data points correspond to an average of 3-15 assessments. As was mentioned, the uncoiling force is predicted to be elongation-independent for velocities below a given corner velocity (under steady-state conditions) but has a logarithmic dependence on elongation speed for forces above (dynamic conditions). The red dashed curves represent fits of the Eqs (2) and (3) to the data, representing steady-state and dynamic conditions, respectively. The rightmost one, which represents the dynamic conditions, has been fitted to 16 sets of velocities, ranging from 1 to 243 µm/s. As can be seen from the plot, the experimental data is in good agreement with the model. The corner velocity for F1C was found to be 1400 nm/s. The slope of the dynamic part of the plot in figure 3b, which is given by $kT / \Delta x_{AT}$, provides one means to obtain a value of the bond length of the LL-bond, $\Delta x_{AT}$.

*4.4. Characterization of F1C pili under temporal conditions*

Temporal behavior can be assessed by relaxation experiments in which the force response in region II, following a sudden halt of elongation in the dynamic regime, is monitored over time. A typical such curve is given in figure 3c. As the elongation is started (denoted "1" in the figure), the force response in region II increases from its steady-state value to that given by the dynamic response corresponding to the particular elongation speed used (as displayed in the Figs 3a and 3b). As the elongation is suddenly halted (at "2"), the force decreases from that given by the dynamic conditions to that corresponding to steady-state (at "3"). Fits of Eq. (6) to such relaxation curves reveal that the decay from a dynamic response to the steady-state level for F1C pili is fast, with a relaxation time $\tau$ (represented by the transition time from 90% to 10% of the difference in force levels) of 0.045 s (since the decay is logarithmic and not exponential, it is not possible to assign an ordinary decay time of the type $\exp(-t/\tau)$ to the relaxation). This provides in turn values for the thermal bond opening rate and the length of the LL-bond, i.e. $k_{AB}^{th}$ and $\Delta x_{AT}$, of $17.5 \pm 1.9$ Hz and $0.38 \pm 0.04$ nm, respectively. These values agree with an evaluation of the data from the DFS measurements, given by the intercept of the data with the x-axis and the slope of the rightmost fit in figure 3b, respectively.

It can be argued that neither the dynamic response, nor the relaxation measurements, is significantly affected by the pilus-bead interaction, the compliance of the bacterial body, or the anchoring to the bacterial body. Since the pili are attached non-specifically to the bead, and such attachment can be considered to be significantly stiffer than any specific single-bond interaction, it is implausible that the attachment to the bead will influence any of the measurements. Moreover, pili are firmly attached to the bacterial cell wall by an anchoring that can maintain several hundreds of pN with no evidence of any compliance. Moreover, the cell wall has been found to be show little elasticity, evidence by the fact that the steady-state compliance of shredded pili is the same as those bound to a bacterium (Jass et al. 2004). This implies that it is not probable that neither the bacterial body, nor the anchoring of the pili to the bacterial cell, will affect the measurements significantly. This is further supported by the fact that this study reveals unusually fast kinetics. The fact that previous investigations of the other pili, predominantly type 1, expressed by the same strain of bacterium, showed orders of magnitude slower kinetics (Andersson et al. 2007) (see discussion below) evidences that the measured dynamic and temporal responses do not originate from the bacterial body or the pili anchoring. In conclusion, this implies that we consider the dynamics effects assessed in this work to primarily originate from the shaft of the pili, primarily its quaternary structure.

## 5. Discussion

The different types of pili expressed by UPEC bacterial cells (type 1, P, S, and F1C) are constructed by different major subunits; FimA, PapA, SfaA$_{II}$/SfaA$_I$, and Foc, respectively (Knight et al. 2002) and assumed to be expressed predominantly at dissimilar parts of the urine tract (UT). Although their quaternary structure shares a common overall helix-like architecture, it has been found that the various pili have dissimilar biomechanical properties. It is thereby of interest to assess these differences and determine to which extent they can be correlated to the preferred region of the UT. The recent assessments of the biomechanical properties of various types of pili by FMOT (Andersson et al. 2006a, c; Andersson et al. 2007; Castelain et al. 2009b; Fällman et al. 2005; Jass et al. 2004), including the present assessment of F1C, have made such a comparison possible. The values of the most relevant biomechanical parameters for the various pili are summarized in Table 1. A comparison of the single-pilus response of F1C with those of other pili shows that qualitatively they have a similar type of behavior; they all elongate in three regions of the same type, I, II, and III, as was illustrated by the grey curve in figure 3a. Quantitatively, however, as is shown in Table 1, the properties of the various pili differ, sometimes quite markedly.

The steady-state uncoiling force for F1C of 26 pN that was found in this work is similar to those of P pili (Andersson et al. 2006a, c; Fällman et al. 2005; Jass et al. 2004) and $S_{II}$ (Castelain et al. 2009b), which have been found to be 28 and 26 pN, respectively, but larger than that of $S_I$ pili and smaller than that of type 1, which have been assessed to 21 pN (Castelain et al. 2009b) and 30 pN (Andersson et al. 2007), respectively. The bond opening length of F1C, $\Delta x_{AB}$, which was found to be 5.6 nm, is virtually identical to that of S pili (5.6 and 5.7 nm for $S_I$ and $S_{II}$, respectively) but larger than that of P pili (3.5 nm). A long bond opening length implies, among other things, that the elongation length in region II (which is given by the product of the bond opening length and the number of subunits in the rod) is large. The bond energy of F1C pili, $\Delta V_{AB}$, which was assessed to $36 \pm 2$ $kT$, was found to be virtually identical to that of type 1 and $S_{II}$, $37 \pm 2$ and $37 \pm 5$ $kT$, respectively, but significantly higher than that of P pili and $S_I$, which have been found to be $24 \pm 1$ and $28 \pm 5$ $kT$, respectively. According to Eq. (2), pili with a large bond opening length would normally have a low steady-state uncoiling force. However, a large bond opening length can be negated by a high bond energy. As can be seen from Table 1, this is the case for F1C as well as $S_{II}$ pili, which have both large bond opening lengths (5.6 and 5.7 nm, respectively) and high bond energies (36 and 37 $kT$, respectively), giving them a steady-state unfolding force (26 pN) that is similar to that of P pili (28 pN), which has a significantly shorter bond opening length (3.5 nm) and lower bond energy (24 $kT$).

Although the steady-state parameter values differ among the various pili to a certain extent, they are still not widely dissimilar. The dynamic and temporal parameters, on the other hand, which govern the elongation speed at a given force exposure and the response behavior under transient conditions, show larger spread. A comparison of the uncoiling-force-*vs.*-elongation-speed-dependence of the various pili reveals that F1C has the highest corner velocity, 1400 nm/s, markedly larger than those of P, $S_{II}$, and $S_I$, which are 400, 450, and 700 nm/s, respectively, and significantly larger than that of type 1, which has been assessed to 6 nm/s. This implies, first of all, that for the lowest velocities probed (which are typically in the order of 0.1 µm/s), the quaternary structure uncoils under steady-state conditions for all types of pili except type 1. For elongation speeds at around a few µm/s and above (which are velocities that are clearly addressable using typical force measuring instrumentation), all pili exhibit their dynamic response.

Secondly, and more importantly, as was shown by Eq. (5), a large corner velocity also implies that a pilus that is exposed to a force above its steady-state uncoiling force will elongate rapidly. This is a drawback for a bacterium that is exposed to forces over an extended amount of time (several seconds), since then the pili might risk to elongate into region III during the force exposure. It has been shown by Björnham and Axner (Björnham and Axner 2009; Björnham and Axner 2010) that a bacterium experiences the longest adhesion lifetime when all adhering pili are simultaneously elongating in region II (thus with all pili experiencing the same force, given by the total force divided by the number of elongating pili); as soon as one pilus reaches region III, the elongation will halt, whereby the force will be unevenly distributed among the various pili (higher for those that are in region III than those that still are in region II). Since the opening rate of a single bond depends exponentially on the force, an uneven force distribution will decrease the adhesion lifetime of the bacterium (Björnham and Axner 2009; Björnham and Axner 2010).

However, a large corner velocity is not necessarily a drawback for a pilus that is exposed to force for only a limited amount of time (shorter than the ratio of the maximum elongation of region II and the elongation speed); conversely, a large corner velocity is more or less imperative if the pilus also would express a fast temporal response (which allows for an ability to handle rapidly fluctuating forces in an efficient manner). The reason is, as can be seen from Eq. (6), that a fast temporal response requires a large thermal extension speed, $\dot{L}^{th}$, which, in turn, according to Eq. (4), demands a large corner velocity, $\dot{L}^*$. Hence, in order to benefit from a fast temporal response, the pili need to exhibit a large corner velocity.

As an illustration of this, it was found that F1C does not only have the largest corner velocity of all pili investigated, it also has the fastest relaxation time. The relaxation time was found to be 0.045 s for F1C, which is a few times shorter than those of P and S pili (which ranges from 0.12 to 0.32 s) and more than two orders of magnitude shorter than that of type 1 (which is 6.7 s). This is obtained by a combination of a fast thermal bond opening rate, $k_{AB}^{th}$, $17.5 \pm 1.9$ Hz, and a short bond length, $\Delta x_{AT}$, $0.38 \pm 0.04$ nm. The thermal bond opening rate is clearly faster than that of S and P pili (which have been found to be 8.8, 1.3, and 0.8 Hz for $S_I$, $S_{II}$, and P, respectively) and distinctly larger than that of type 1, which is 0.016 Hz, whereas the bond length is shorter than those of the others (0.56, 0.66, 0.76, and 0.59 Hz for $S_I$, $S_{II}$, P, and type 1, respectively). All this implies that although F1C pili cannot sustain large forces for more than a limited amount of time, they are expected to adjust rapidly to fluctuating forces.

The dissimilarity in the dynamic and temporal responses of the major types of pili expressed by UPEC, i.e. P, type 1, S and F1C, can hypothetically be correlated to the environmental conditions in different parts of the UT. For example, it is known that the urine is transported in boluses via ureters by a peristaltic activity from the kidney to the bladder (Griffiths 1987; Griffiths et al. 1987), and that the bladder expels urine via the urethra (Ozawa et al. 1998). The flow in the boluses exposes the bacteria to widely fluctuating forces, of both shear (including reversal wall shear forces) and normal (i.e. perpendicular) type (Vogel et al. 2004), similar to those in the ileum (Jeffrey et al. 2003). It can be assumed that in order to maintain firm adhesion under such rapidly changing conditions the pili should preferably have a fast kinetics. Hence, the measured data suggest that F1C is a type of pili that predominantly is expressed in the upper UT.

## 6. Conclusions

In conclusion, force measuring optical tweezers, FMOT, have proven to be a suitable tool for studies of bacterial pili with respect to their biomechanical properties. The technique allows for a scrutiny of the biomechanical parameters of the overall fimbrial architecture on a single pilus level and can therefore directly assess properties related to their uncoiling/recoiling when exposed to force. As has been discussed recently, the particular force-*vs.*-elongation response of this helix-like structure suggests that piliated UPEC bacteria are able to share an external shear force applied by urine flow among several pili in such a way that the force is fairly evenly distributed among the various pili, which helps the bacterium to maintain adhesion (Björnham and Axner 2009).

In order to add to the library of biomechanical features of pili that has been assembled during the latter years (Andersson et al. 2006a, c; Andersson et al. 2007; Castelain et al. 2009b; Fällman et al. 2005; Jass et al. 2004), and to correlate experimental findings to various hypotheses, this work has focused on F1C pili, which is a type of pili that so far has not been studied in terms of force spectroscopy, and assessed its properties under steady-state, dynamic, as well as temporal conditions. It was specifically found that although the steady-state behavior of F1C pili resembles those of the other types (P, type 1, and S pili) their dynamic and temporal properties differ; F1C pili have a faster kinetics and a shorter relaxation time than all other pili so far studied. It is most plausible that the different properties of the various types of pili reflect the local environment in which they predominantly are being expressed (Andersson et al. 2007). It is therefore of interest to correlate the various properties of pili and the adhesion properties of bacteria expressing such pili to the environmental conditions in various parts of the urinary tract. However, although a first and simple correlation was made in this work, which indicates that F1C predominantly is a type of pili that are supposed to be expressed in the upper urinary tract, where the most fluctuating force conditions are assumed to be found, a detailed correlation needs to consider a multitude of aspects, and we will aim at this in future work. Moreover, using the models previously derived for the elongation and retraction of pili in the presence of force, and the parameter values currently being assessed for the various pili, the response of various types of bacteria adhering by multipili adhesion can presumably soon be predicted (e.g. by simulations) under a variety of conditions, including those that are non-trivial to assess experimentally. Characterizations of the biomechanical properties of helix-like pili might therefore open up new doors to the understanding of how piliated bacteria behave under *in vivo* conditions and can hopefully also shed light onto their prevalence in the early stages of infection caused by UPEC bacteria.

### Acknowledgements

The authors are thankful to Monica Persson for her technical assistance with strains/plasmids. This work was performed within the Umeå Centre for Microbial Research (UCMR) and was carried out in the frame of the European Virtual Institute for Functional Genomics of Bacterial Pathogens (CEE LSHB-CT-2005-512061) and the ERA-NET project "Deciphering the intersection of commensal and extraintestinal pathogenic *E. coli*". We acknowledge economical support from the Swedish Research Council (621-2008-3280), from the Kempe foundations, and from Magnus Bergvall's foundation.

**Figures captions**

**Figure 1**: AFM micrograph of HB101/pBSN50 cells expressing F1C pili, scale bar, 2 µm.

**Figure 2**: (a) Typical force-*vs.*-elongation curve when a single pilus is uncoiling (black curve) and recoiling (grey curve) under steady-state conditions at 0.1 µm/s. The characteristic elongation regions I, II and III are marked in the figure. Red curves represent the multipili pili interactions during the uncoiling process. The region II$^{(n)}$ of $n$ pili uncoiling is $n$-fold the region II$^{(1)}$ of a single pilus, as it is for region II$^{(3)}$ and II$^{(2)}$ for 3 and 2 pili, respectively. Black arrows indicate typical events when one pilus is detaching from the trapped bead, reducing thereby the number of uncoiling pili attached to the bead. (b) Distribution of the uncoiling force of single pili over n=230 measurements from independent experiments, with bacteria from different cultures, conducted over a period of about one year.

**Figure 3**: Force-*vs.*-elongation curve at 0.1 µm/s, i.e., under steady-state conditions (black and grey). The other colored curves represent measurements that were performed at different elongation velocity (3, 9 and 64 µm/s) under dynamic conditions, over region III on the same pilus. (b) DFS curve, i.e., uncoiling force-*vs.*-elongation velocity $\dot{L}$ (lin-log plot). Each data are an average of about 3-15 measurement with their standard deviation (Y-error bars). The intersection of the two linear fits gives the corner velocity $\dot{L}^*$ representing the boundary between the steady-state ($\dot{L} < \dot{L}^*$) and the dynamic regimes ($\dot{L} > \dot{L}^*$). (c) Relaxation curve, i.e. force-*vs.*-time, smoothened (orange) and fitted with Eq. (6) (dashed red). The stage is translating at a set velocity

from time 1 to time 2 (arrows 1 and 2). The force raises the corresponding uncoiling force level over a given distance (typically 1.5-3 µm). Once the stage is halted at time 2 (arrow 2), the force decreases to time 3 (arrow 3) reaching the uncoiling force level under steady-state regime as it is before time 1.

**Figures**

Figure 1, one column

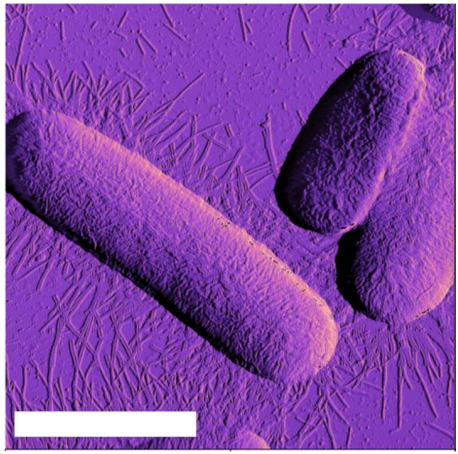

Figure 2, over two columns

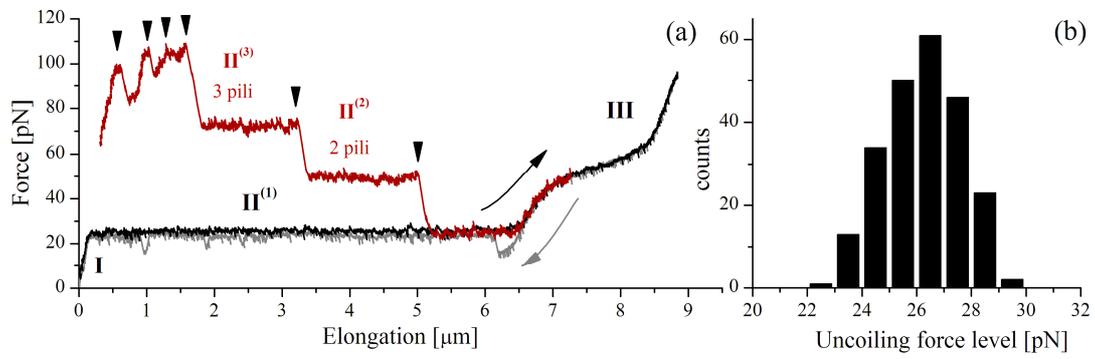

Figure 3, over one column

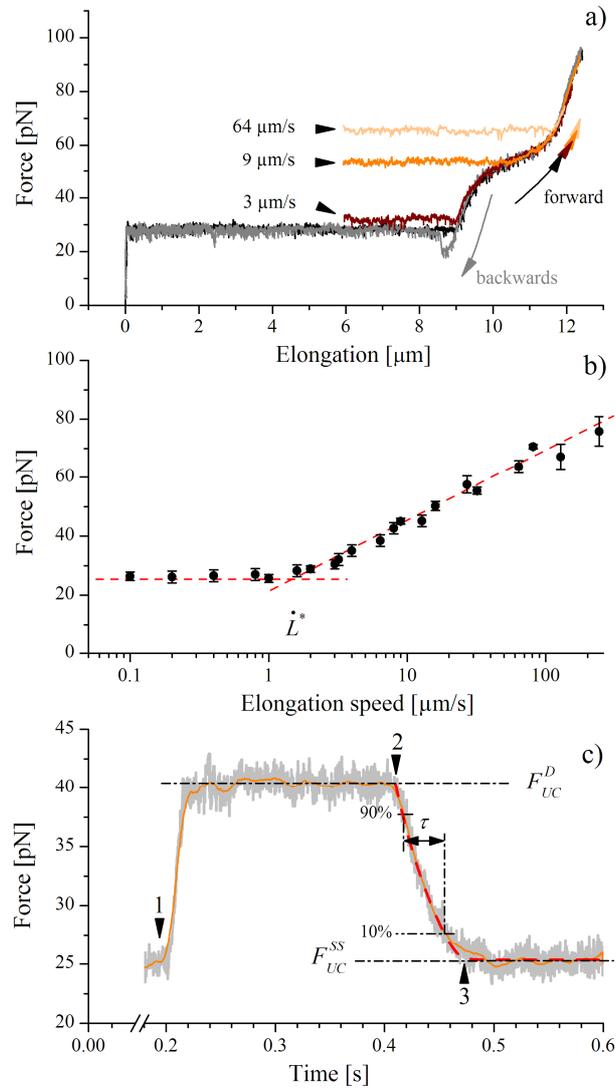

**Table 1**: Summary of assessed parameters. The data for $S_I/S_{II}$, P and Type 1 are taken from (Castelain et al. 2009b), (Andersson et al. 2006c) and (Andersson et al. 2007), respectively.

| Strain/ | HB101 | | | | |
|---|---|---|---|---|---|
| Plasmid/ | pBSN50 | pANN801-13 | pAZZ50 | pPAP5 | pPKL4 |
| Pili/ | F1C | $S_I$ | $S_{II}$ | P | Type 1 |
| $F_{UF}$ (pN) | 26 ± 1 | 21 ± 2 | 26 ± 1 | 28 ± 2 | 30 ± 2 |
| $\Delta x_{AB}$ (nm) | 5.6 ± 0.4 | 5.6 ± 0.5 | 5.7 ± 0.5 | 3.5 ± 0.1 | 5.0 ± 0.3 |
| $\Delta V_{AB}$ ($kT$) | 36 ± 2 | 28 ± 5 | 37 ± 5 | 24 ± 1 | 37 ± 2 |
| $\dot{L}^{*}$ (nm/s) | 1400 ± 160 | 700 ± 100 | 450 ± 150 | 400 ± 100 | 6 ± 3 |
| $\tau$ (s) | 0.045 | 0.18 | 0.32 | 0.12 | 6.7 |
| $k_{AB}^{th}$ (Hz) | 17.5 ± 1.9 | 8.8 ± 5.3 | 1.3 ± 0.9 | 0.8 ± 0.5 | 0.016 ± 0.009 |
| $\Delta x_{AT}$ (nm) | 0.38 ± 0.04 | 0.56 ± 0.14 | 0.66 ± 0.08 | 0.76 ± 0.11 | 0.59 ± 0.06 |